\documentclass[apjl]{emulateapj}
\usepackage{natbib}
\usepackage{ graphicx}
\usepackage{ subfigure}
\usepackage{multirow}

\shorttitle{Growth of Early Supermassive BHs}
\shortauthors{DeGraf et al.}

\begin{document}

\title{Growth of Early Supermassive Black Holes and the High-Redshift Eddington Ratio Distribution}
\author{C. DeGraf, \altaffilmark{1} T. Di Matteo, \altaffilmark{1} N. Khandai,
\altaffilmark{1} R. Croft \altaffilmark{1}}
\altaffiltext{1} {McWilliams Center for
       Cosmology, Carnegie Mellon University, 5000 Forbes Avenue, Pittsburgh,
       PA 15213, USA}

\def\simgt{\lower.5ex\hbox{$\; \buildrel > \over \sim \;$}}
\newcommand{\chimps}{$h^{-1}$Mpc }
\newcommand{\hisms}{$h^{-1}$M$_{\sun}$}
\newcommand{\omegaM}{$\Omega_{\rm m}$}
\newcommand{\omegaB}{$\Omega_{\rm b}$}
\newcommand{\omegaL}{$\Omega_{\rm \Lambda}$}
\newcommand{\sigmaeight}{$\sigma_{8}$}

\begin{abstract}

Using a new large-scale $(\sim 0.75 \: Gpc)^3$ hydrodynamic cosmological 
simulation we
investigate the growth rate of supermassive black holes in the early 
universe
($z \simgt 4.75$).  Remarkably we
find a clear peak in the typical Eddington ratio ($\lambda$) at black 
hole
masses of $ 4-8 \times 10^{7} M_\odot$ (typically found in halos of $\sim 
7
\times 10^{11}-1 \times 10^{12} M_\odot$), independent of redshift and
indicative that most of BH growth occurs in the cold-flow dominated 
regime.
Black hole growth is by and large regulated by the evolution of gas density.
The typical Eddington ratio at a given mass scales simply as cosmological
density $(1+z)^3$ and the peak  is caused by the competition between
increased gas density available in more massive hosts, and a decrease due 
to
strong AGN feedback that deprives the black hole of sufficient gas to fuel
further rapid growth in the high mass end.
In addition to evolution in the mean Eddington ratio, we show that the
distribution of $\lambda$ among both mass-selected and luminosity-selected
samples is approximately log-normal.
We combine these findings into a single log-normal fitting formula for the 
distribution of
Eddington ratios as a function of $(M_{\rm{BH}},z)$.  This formula can be 
used
in analytic and semi-analytic models for evolving black hole populations,
predicting black hole masses of observed quasars, and, in conjunction with
the observed distribution of Eddington ratios, can be used to constrain 
the
black hole mass function.

\end{abstract}

%\begin{keywords}
%\end{keywords}

\section{Introduction}
\label{sec:Introduction}

It has been well established that supermassive black holes are present in the
center of most galaxies \citep{KormendyRichstone1995}, and that they are
correlated with the properties of their hosts \citep{Magorrian1998, FerrareseMerritt2000,
  Gebhardt2000, Tremaine2002, GrahamDriver2007}.  These correlations provide
strong evidence that the growth of a black hole and the evolution of its host
galaxy directly influence one another, such that black hole growth is a
important aspect of understanding galactic evolution and vice versa.

In general, the link between black hole and galactic evolution is attributed
to some form of quasar feedback \citep{BurkertSilk2001, Sazonov2004, Springel2005, Churazov2005,
  DiMatteo2005, Bower2006,
  CiottiOstriker2007, Sijacki2007, Hopkins2007} which can
result in the self-regulation of the growth of the black hole \citep[see,
  e.g.][]{DiMatteo2005}.  In this model we would expect black holes to grow
rapidly during their early lifetime (i.e. while at low mass), until some
point at which the black hole feedback begins to significantly affect its
environment, resulting in a noticeable decline in growth rate.  This effect has been
observed in individual black hole histories,
but such investigations  \citep[see, e.g.][]{Sijacki2009, DiMatteo2011} have tended to focus on the largest mass black holes,
primarily to explain how black holes could grow rapidly enough to produce the
extremely large masses ($\sim 10^9
M_\odot$ by $z \sim 6$) found in 
observations by the Sloan Digital Sky Survey \citep[e.g.][]{Fan2006,Jiang2009}.  In this paper we take advantage of a new, very large simulation to investigate
the growth histories of early universe black holes across a wide range of
masses, probing both the mean and the distribution of growth rates for
black holes across a wide range of masses and luminosities, and provide fits
for these distributions.

\section{Method}
\label{sec:Method}

In this paper we use a new cosmological hydrodynamic
simulation of a $533 \: h^{-1}$ Mpc box specifically
intended for high-redshift investigations.  The simulation uses the massively
parallel cosmolocial TreePM-SPH code P-GADGET \citep[an updated version of
  GADGET-2, see][]{2005MNRAS.364.1105S} incorporating a multi-phase ISM model
with star formation \citep{SpringelHernquist2003} and black hole accretion and feedback
\citep{SpringelFeedback2005, DiMatteo2005}, has a gravitational softening
length of $5 \: h^{-1}$ kpc and mass resolution of $2.8 \times
10^8 M_\odot$ for dark matter and $5.7 \times 10^7 M_\odot$ for gas.  

Within the simulation, black holes are modeled as collisionless sink particles
which form in newly emerging and resolved dark matter halos.  These halos are
found by calling a friends of friends group finder at regular intervals (in
time intervals spaced by $\Delta \log a = \log 1.25$).  
Any group above a threshold mass of $5 \times 10^{10}
h^{-1} M_\odot$ not already containing a black hole is provided one by
converting its densest particle to a sink particle with a seed mass of
$M_{\rm{BH,seed}} = 5 \times 10^5 h^{-1} M_\odot$.  This seeding prescription
is chosen to reasonably match the expected formation of supermassive black
holes by gas directly collapsing to BHs with $M_{\rm{BH}} \sim
M_{\rm{seed}}$ \citep[e.g.][]{BrommLoeb2003, Begelman2006} or by PopIII stars
collapsing to $\sim 10^2 M_\odot$ BHs at $z \sim 30$ \citep{Bromm2004, Yoshida2006} followed by sufficient
exponential growth to reach $M_{\rm{seed}}$ by the time the host halo reaches
$\sim 10^{10} M_\odot$.  Following insertion, BHs grow in mass by accretion of
surrounding gas and by merging with other black holes.  Gas is accreted
according to $\dot{M}_{\rm BH} = \alpha \frac
  {4 \pi G^2 M_{\rm BH}^2 \rho_{\rm{BH}}}{(c_s^2 + v^2)^{3/2}}$, where
  $\rho_{\rm{BH}}$ is the local gas density (determined from the gas particles
  within the black hole kernel), $c_s$ is the local sound speed, $v$ is
  the velocity of the BH relative to the surrounding gas, and $\alpha$ is introduced to correct for the reduction of the gas density close
  to the BH due to our effective sub-resolution model for the ISM. To allow for the
  initial rapid BH growth necessary to produce sufficiently massive BHs at
  early time ($\sim 10^9 M_\odot$ by $z \sim 6$) we allow for mildly
  super-Eddington accretion, but limit it to a maximum of $3 \times \dot{M}_{\rm{Edd}}$
  to prevent artificially high values.

The BH is assumed to radiate with a bolometric luminosity proportional to the
accretion rate, $L = \eta \dot{M}_{\rm{BH}} c^2$ \citep{ShakuraSunyaev1973},
where the radiative efficiency $\eta$ is fixed to 0.1 throughout the
simulation and our analysis.  To model the expected coupling between the
liberated radiation and the surrounding gas, 5 per cent of the luminosity is
isotropically deposited to the local black hole kernel as thermal energy.  The
5 per cent value for the coupling factor is based on galaxy merger simulations
such that the normalization of the $M_{\rm{BH}}-\sigma$ relation is reproduced
\citep{DiMatteo2005}.  

The second mode of black hole growth is through mergers which occur when dark
matter halos merge into a single halo, such that their black holes fall toward the
center of the new halo, eventually merging with one another.  In cosmological
volumes, it is not possible to directly model the physics of the infalling BHs
at the smallest scales, so a sub-resolution model is used.  Since the mergers
typically occur at the center of a galaxy (i.e. a gas-rich environment), we assume the final coalescence
will be rapid \citep[e.g.][]{Mayer2007}, so we merge the
BHs once they are within the spatial resolution of the simulation.  However,
to prevent merging of BHs which are rapidly passing one another, mergers are
prevented if the BHs' velocity relative to one another is too high (comparable
to the local sound speed).  

The model used for black
hole creation, accretion and feedback has been investigated and discussed in
\citet{Sijacki2007, DiMatteo2008, Colberg2008, Sijacki2009,
    DeGraf2010, DeGrafClustering2010}, finding it does a good job reproducing
the $M_{\rm{BH}}-\sigma$ relation, the total black hole mass density
\citep{DiMatteo2008}, the QLF \citep{DeGraf2010}, and the expected black hole
clustering behavior \citep{DeGrafClustering2010}.  This simple model thus appears
to model the growth, activity, and evolution of supermassive black holes in
a cosmological context surprisingly well (though the detailed treatment of the accretion
physics is infeasible for cosmological scale simulations).  We also note that
\citet{BoothSchaye2009} and \citet{Johansson2008} have adopted a very similar model,
and have independently investigated the parameter space of the reference model
of \citet{DiMatteo2008}, as well as varying some of the underlying
prescriptions.  In addition, this simulation has previously been used to
investigate the growth of the first very massive black holes
\citep{DiMatteo2011}, statistical properties of quasars
\citep{DeGraf2011EarlyBHs}, and large scale high-resolution imaging \citep{Feng2011}.  For further details on the simulation methods and
convergence studies done for similar simulations, see \citet{DiMatteo2008}.  

Because the simulation saves the complete set of black hole properties (mass,
accretion rate, position, local gas density, sound speed, velocity, and BH
velocity relative to local gas) for each BH at every timestep, the black hole output for
such a large simulation is prohibatively difficult to analyze using previous
techniques.  For this reason, \citet{Lopez2011} developed a relational database
management system specifically for this simulation.  A similar strategy has
also been followed in the analysis of the Millenium simulation
\citep{Lemson2006}.  In
addition to providing a substantially more efficient query system for
extracting information, this database is significantly more flexible than
traditional approaches.  For a complete summary of the database format and its efficiency,
please see \citet{Lopez2011}.

\section{Results}
\label{sec:Results}

\subsection{Typical Black Hole Growth Rates}
\label{sec:growth}

\begin{figure}
\centering
\includegraphics[width=9cm]{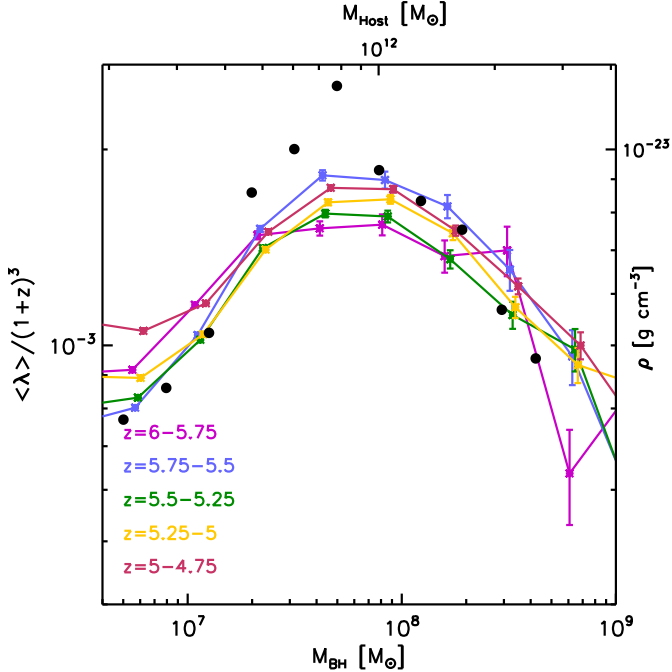}
\caption{\textit{Colored lines:} The mean Eddington ratio ($\langle \lambda \rangle$) as a function of $M_{\rm{BH}}$ for
  several redshift ranges, scaled by $\frac{1}{\left ( 1+z \right)^3}$, with
  Poisson error bars [Note that the datapoints' x-positions for each z-bin have been
  shifted to the right (3\% increase for each z-bin) such that the error
  bars are distinguishable].  We also show the typical host halo mass
  corresponding to the given BH mass on the top axis.  \textit{Filled circles:}
  Average gas density at the BHs position ($\rho_{\rm{BH}}$) for $z=4.75-5$.
}
\label{meangrowth}
\end{figure}

To quantify the growth rate of black holes, we use the mean Eddington ratio
($\lambda = \frac{\dot{M}_{\rm{BH}}}{\dot{M}_{\rm{edd}}}$) which we
calculate for each black hole over a finite time interval.  Because we have
the complete BH growth history, we are able to compute this quantity based
solely on the gas accretion, and neglect any mass gained through black hole
mergers (though we find the mass gained by mergers to be small enough to have a negligible
effect on our results).  In Figure
\ref{meangrowth} we plot $\langle \lambda\rangle/(1+z)^3$ as a function
  of $M_{\rm{BH, initial}}$ for several redshift ranges.  We plot $\langle
  \lambda \rangle/(1+z)^3$ rather than $\langle \lambda \rangle$ for
  two reasons: First, to
  show that the dependence of $\langle \lambda \rangle$ on $M_{\rm{BH}}$
  is independent of redshift (at least for $z \ge 4.75$), and second to show that $\langle
  \lambda \rangle \propto (1+z)^3$. 

  Regardless of redshift considered, we find similar behavior for Eddington ratio
  with respect to mass: more massive black holes grow faster than low mass
  black holes up to a peak growth rate at $M_{\rm{BH}} \sim 4-8 \times 10^7
  M_{\odot}$, while the black holes above this characteristic mass grow more slowly.  Thus black holes grow
  fastest (relative to their current mass) while at intermediate
  masses, and grow slower at higher mass.

\begin{figure}
\centering
\includegraphics[width=8cm]{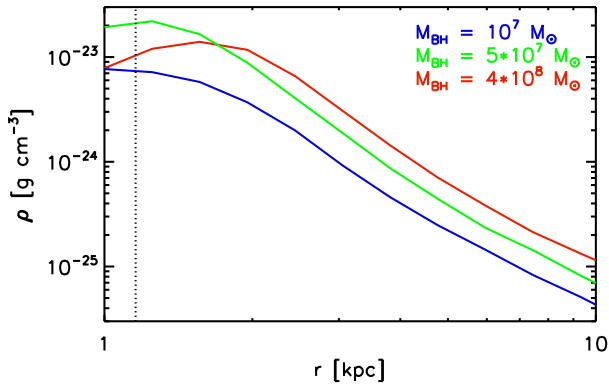}
\caption{  Gas density profiles averaged among 100 black holes with mass
  $\sim 10^7 M_\odot$ (blue), $\sim 5 \times 10^7 M_\odot$ (green), $\sim 4
  \times 10^8 M_\odot$ (red).  Dotted line shows the gravitational softening
  length.}
\label{gasprofile}
\end{figure}

  We find this peak in the Eddington ratio to be caused by the change in the
  local gas density available for fueling BH growth.  We plot the evolution in
  local gas density ($\rho_{\rm{BH}}$, the density of gas contributing to $\dot{M}_{\rm{BH}}$) with
  mass in Figure \ref{meangrowth}, showing a clear peak at $\sim 5 \times 10^7
  M_\odot$.  We note that neither the sound speed nor the BH velocity (the other factors in the calculation of
  $\dot{M}_{\rm{BH}}$) exhibit a peak with respect to $M_{\rm{BH}}$,
  confirming that the peak Eddington ratio is caused by the evolution
  in the local gas density.  To show how the gas density evolves, in 
  Figure \ref{gasprofile} we show the gas density profiles around BHs
  below the Eddington ratio peak ($\sim 10^7 M_\odot$ - blue), at the peak
  ($\sim 5 \times 10^7 M_\odot$ - green), and above the peak ($\sim 4 \times
  10^8 M_\odot$ -- red), each averaged across 100 BHs.  In general we find the
  gas density profile to grow with
  $M_{\rm{BH}}$ until $M_{\rm{BH}} \sim 5 \times 10^7 M_\odot$ (as expected
  for BHs found in more massive halos).  Above $\sim 5
  \times 10^7 M_\odot$ the gas density away from the BH continues to grow, but
  the innermost density is suppressed, with the suppression growing with $M_{\rm{BH}}$
  in both magnitude and distance.  This suppression of the local gas density is caused by the
  feedback of the black hole, with the stronger feedback of high-mass BHs
  producing the strongest effect (see \citet{DiMatteo2011} for detailed investigation of
  feedback among massive BHs).

  We also show the typical mass of halos hosting a given $M_{\rm{BH}}$ on the
  top axis, noting that the Eddington ratio peaks at a host halo mass of $\sim 7 \times 10^{11}-1 \times 10^{12} M_\odot$.
  This mass very closely matches the critical shock heating scale of $\sim 6
  \times 10^{11} M_\odot$ \citep[][and consistent with our simulation]{DekelBirnboim2006, Dekel2009}, above which
  infalling gas is shock heated near the virial radius to the virial
  temperature of the halo.  \citet{DekelBirnboim2006} suggest that in these
  halos AGN feedback becomes more significant, since the dilute shock-heated
  gas will be more susceptible to heating and pushing by the central AGN.
  This would thus produce a suppression in the gas density profile, consistent
  with the picture described above and the downturn in Figure \ref{meangrowth}.

\begin{figure}
\centering
\includegraphics[width=9cm]{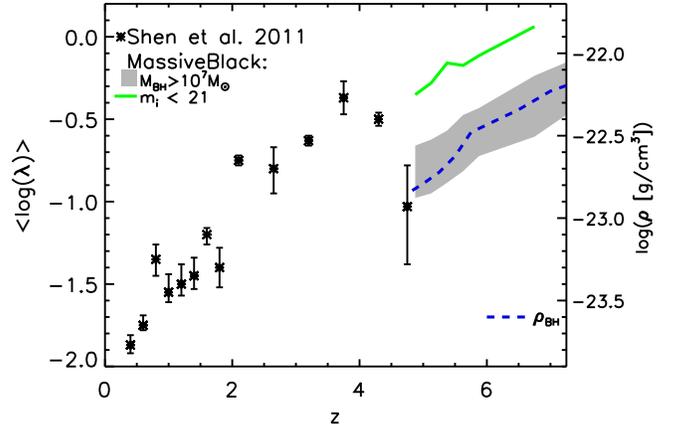}
\caption{Redshift evolution of the Eddington ratio for black holes with
  $M_{\rm{BH}} > 10^7 M_\odot$ (shaded region shows 1-$\sigma$ standard deviation in
  log($\lambda$)) and i-band magnitude $m_i < 21$ (green line) compared with data from
  \citet{Shen2011} (black asterisks). We
  also show the evolution in the gas density around BHs for comparison (blue
  dashed line). }
\label{zevolution}
\end{figure}

In addition to the evolution in $\lambda$ with $M_{\rm{BH}}$, Figure
\ref{meangrowth} also shows that $\lambda$ evolves with redshift as $\sim
\left (1+z \right)^3$, which is also caused by the evolution in the
local gas density.  In Figure \ref{zevolution} we show the evolution in
$\langle \rm{log} \left ( \lambda \right ) \rangle$ with redshift among
$M_{\rm{BH}} > 10^7 M_\odot$ BHs (shaded region, showing 1-$\sigma$ standard deviation).  We plot the average gas
density at the BH (blue dashed line), showing the evolution in $\lambda$ is
primarily caused by the evolution in $\rho_{\rm{BH}}$ (recall $\dot{M}_{\rm{BH}} \propto
\rho_{\rm{BH}}$).  We also compare to 
observational measurements of \citet{Shen2011} (black asterisks), showing that
this general redshift evolution is consistent with current observations, and
the normalization is approximately consistent if we use a similar magnitude
cut (i-band magnitude
$m_i < 21$ - green line).

\subsection{Eddington Ratio Distributions}
\label{sec:model}

\begin{figure}
\centering
\includegraphics[width=9cm]{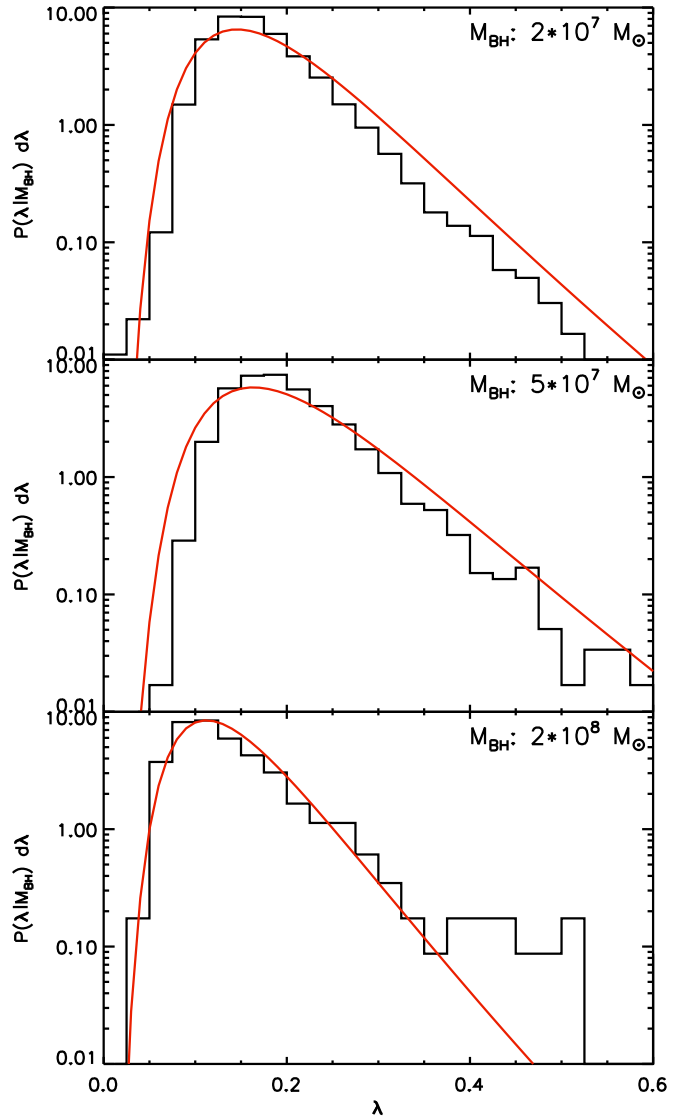}
\caption{Eddington ratio distribution for black holes at three different mass
  scales (black) and the predicted distribution from Equations
  \ref{eqn:probability}\&\ref{eqn:mu_m} (red curves).}
\label{lambdadist}
\end{figure}

In addition to investigating the mean Eddington ratio, we also study the
distribution of $\lambda$ among comparable BHs.
Previous work on the $\lambda$-distribution has often found roughly log-normal
distributions using both observational \citep{Kollmeier2006, Netzer2007,
  NetzerTrakhtenbrot2007, Willott2010, Trakhtenbrot2011} and phenomenological
approaches \citep{Shankar2011} [though \citet{Aird2011} find $\lambda$ to follow a power law when
selected for host stellar mass, rather than BH mass].  However, these
observational studies necessarily
incorporate several uncertainties, such as sample selection and scatter in
black hole mass estimators, which we can bypass, using our simulation to
probe our black holes' Eddington ratios directly.
In Figure \ref{lambdadist} we show the distribution of Eddington ratios among
black holes selected by $M_{\rm{BH}}$ (black histograms).  We find that the
distribution produced by our simulation is indeed log-normal, in keeping with
 observational findings \citep{Kollmeier2006, Netzer2007,
  NetzerTrakhtenbrot2007, Willott2010, Trakhtenbrot2011}.  In particular, we
note that the distribution remains log-normal regardless of the mass
considered, with Figure \ref{lambdadist} showing this holds among black holes
that are below, at, and above the peak observed in Figure \ref{meangrowth}.  

Because we find $\lambda$ to follow a log-normal distribution and
the mean of that distribution obeys a well-defined curve with $M_{\rm{BH}}$
(Figure \ref{meangrowth}), we are able to provide a general fitting formula
for $P(\lambda|M_{\rm{BH}},z)$, the probability distribution of black hole
Eddington ratios as a function of redshift and black hole mass:
\begin{equation}
  P(\lambda|M_{\rm{BH}},z) = \frac{1}{\lambda \sigma_{m} \sqrt{2 \pi}}
  e^{- \frac{(\rm{ln}(\lambda)-\mu_{m})^2}{2 \sigma_{m}^2}}
\label{eqn:probability}
\end{equation}
where $\mu_{m}$ and $\sigma_{m}$ are the mean and standard deviation of
$\rm{ln}(\lambda)$, respectively, and are fit by $\sigma_m \sim 0.39$ and 
\begin{equation}
\mu_m = (1+z)^3 A e^{-\left (\rm{log}_{10} \left ( \frac{M_{\rm{BH}}}{M_{\mu}}
  \right ) \right)^2/2\sigma_0^2},
\label{eqn:mu_m}
\end{equation}
with $A \sim .00094$, $M_\mu= 5 \times 10^7 M_\odot$, and $\sigma_0 \sim
0.85$.  In Figure \ref{lambdadist} we plot the distribution predicted by
Equations \ref{eqn:probability}\&\ref{eqn:mu_m} (red curve) compared the the actual distribution, showing that
this simple formula is capable of reproducing the distribution of $\lambda$ for BHs in our simulation
across a wide range of masses and redshifts, without requiring knowledge of
individual black hole environments.

\begin{figure}
\centering
\includegraphics[width=9cm]{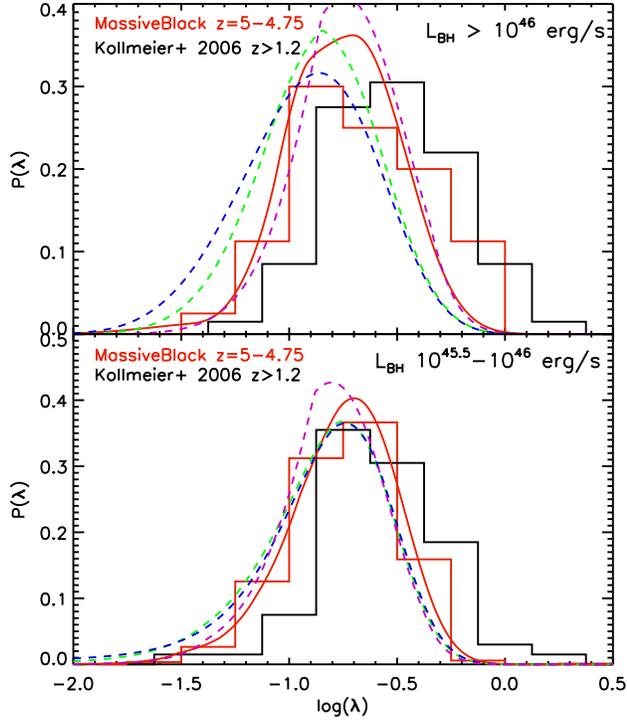}
\caption{Distribution of Eddington ratios for BHs in our simulation (red histogram)
  compared with observational data from \citet{Kollmeier2006} (black histogram) for two
  luminosity bins.  We also show the predicted distribution based on our fitting
  function (Equations \ref{eqn:probability_lum}) using our simulation's
  mass function (solid red), the \citet{Shankar2009} base mass function
  (dashed green),
  the \citet{Shankar2009} mass function derived from the \citet{Hopkins2007}
  luminosity function
  (dashed blue), and the mass function of \citet{2007ApJ...669...45H} (dashed pink). }
\label{Kollmeiercomparison}
\end{figure}

In addition to the distribution for a mass-selected sample, in Figure
\ref{Kollmeiercomparison} we show the Eddington ratio distribution from our
simulation (red histogram) compared to the observed distribution from
\citet{Kollmeier2006} (black histogram) for two luminosity selected samples.  We again
note that the distribution is described by a roughly log-normal distribution,
and that our simulation is approximately consistent with observational
results.

Furthermore, by combining $P(\lambda|M_{\rm{BH}},z)$ with the black hole mass
function ($\Phi_{\rm{BH}}$) we can obtain the Eddington ratio probability
distribution for a luminosity-selected sample:
\begin{equation}
  P(\lambda|L_{\rm{BH}},z)=\frac{\Phi_{\rm{BH}}(M_{\rm{BH}})
  P(\lambda|M_{\rm{BH}},z)}{\int_0^\infty \Phi_{\rm{BH}}(M_{\rm{BH}}) P(\lambda|M_{\rm{BH}},z) d\lambda}
\label{eqn:probability_lum}
\end{equation}
where $M_{\rm{BH}}=\frac{\sigma_T L_{\rm{BH}}}{4 \pi G m_p c \lambda}$.  In
Figure \ref{Kollmeiercomparison} we plot this predicted probability
distribution (using our simulation's mass function) in red, showing
$P(\lambda|L_{\rm{BH}},z)$ is well predicted in this manner.  We note that
this approach is significant as it provides a potentially powerful tool for
constraining the black hole mass function using observations of the Eddington
ratio distribution.  We show this in Figure \ref{Kollmeiercomparison} by
plotting $P(\lambda|L_{\rm{BH}},z)$ based on three different local mass functions:
the \citet{Shankar2009} mass function (dashed green); the \citet{Shankar2009}
mass function derived from the \citet{Hopkins2007} luminosity function (dashed blue), and the
mass function of \citet{2007ApJ...669...45H} (dashed pink).  Because
$P(\lambda|L_{\rm{BH}})$ is sensitive to the slope of $\Phi_{\rm{BH}}$, the distribution
of $\lambda$ at high $L_{\rm{BH}}$ (where the mass function is steepest)
varies substantially with the mass function used, suggesting that with
improved statistics from upcoming surveys, we could use the observed
$P(\lambda|L_{\rm{BH}})$ to constrain the slope of the black hole mass
function at high redshift, even without measurements of the black hole masses.

\section{Conclusions}
\label{sec:Conclusions}

With a new large-scale simulation, we show that the growth of black holes
tends to follow a typical growth pattern.  In general, we find that black
holes grow more rapidly at higher redshift than comparable black holes at
lower redshift, characterized by $\lambda \propto (1+z)^3$.  This scaling is 
caused by the redshift evolution in the gas density about the black holes, and
is comparable to current observational data from \citet{Shen2011}.  

The typical Eddington ratio also scales with $M_{\rm{BH}}$ such that $\lambda$
peaks at $M_{\rm{BH}} \sim 4-8 \times 10^7 M_\odot$ (typically found in halos
of $\sim 7 \times 10^{11}-1 \times 10^{12} M_\odot$).  This peak is caused by
evolution in the density of the gas at halo centers that is available to fuel black hole growth.  In general, more massive
black holes are found in more massive halos with correspondingly higher gas
densities, hence $\lambda$ grows with $M_{\rm{BH}}$ for low masses.  
However, above $M_{\rm{BH}} \sim 5 \times 10^7 M_\odot$ black
hole feedback has a sufficiently strong effect on the local environmen to suppress the
density of the nearby gas.  Thus although these more massive black holes are
found in more massive halos with correspondingly higher gas densities in general, the feedback has significantly lessened the density of the
innermost gas where accretion occurs.  This suppression of the
local gas density is stronger for more massive BHs, and causes
$\lambda$ to decrease for $M_{\rm{BH}} \simgt 5 \times 10^7 M_\odot$.  

Although the local environment is important for the accretion rate of
individual black holes, we show that the distribution of Eddington ratios
follows a roughly log-normal distribution regardless of the black hole
population considered, consistent with current observational findings.  We
use this, together with the evolution in $\langle \lambda \rangle$, to
provide a simple fitting formula for the distribution of Eddington ratio with
$(M_{\rm{BH}},z)$.  This general forumla can be used for predicting the growth/evolution of black hole
populations in theoretical and semi-analytic models (such as the evolution of
the black hole mass function), for predicting the mass of observed
high-redshift quasars, and, in conjunction with upcoming observations of the
$\lambda$-distribution, to constrain the slope of the high-redshift black
hole mass function.

\section*{Acknowledgments}
This work was supported by the National Science Foundation, NSF Petapps,
OCI-0749212 and NSF AST-1009781.  The simulations used in this paper were carried out on Kraken
at the National Institute for Computational Sciences (http://www.nics.tennessee.edu/).

 \bibliographystyle{mn2e}	% or "unsrt", "alpha", "abbrv", etc.
 \bibliography{astrobibl}	% use data in file "astrobibl.bib"

\begin{thebibliography}{50}
\expandafter\ifx\csname natexlab\endcsname\relax\def\natexlab#1{#1}\fi

\bibitem[{Aird} et~al.(2011){Aird}, {Coil}, {Moustakas} et~al.]{Aird2011}
{Aird} J., et~al., 2011, ArXiv e-prints

\bibitem[{Begelman} et~al.(2006){Begelman}, {Volonteri} \&
  {Rees}]{Begelman2006}
{Begelman} M.~C., {Volonteri} M., {Rees} M.~J., 2006, \mnras, 370, 289

\bibitem[{Booth} \& {Schaye}(2009)]{BoothSchaye2009}
{Booth} C.~M., {Schaye} J., 2009, \mnras, 398, 53

\bibitem[{Bower} et~al.(2006){Bower}, {Benson}, {Malbon} et~al.]{Bower2006}
{Bower} R.~G., {Benson} A.~J., {Malbon} R., {Helly} J.~C., {Frenk} C.~S.,
  {Baugh} C.~M., {Cole} S., {Lacey} C.~G., 2006, \mnras, 370, 645

\bibitem[{Bromm} \& {Larson}(2004)]{Bromm2004}
{Bromm} V., {Larson} R.~B., 2004, ARA\&A, 42, 79

\bibitem[{Bromm} \& {Loeb}(2003)]{BrommLoeb2003}
{Bromm} V., {Loeb} A., 2003, \apj, 596, 34

\bibitem[{Burkert} \& {Silk}(2001)]{BurkertSilk2001}
{Burkert} A., {Silk} J., 2001, \apjl, 554, L151

\bibitem[{Churazov} et~al.(2005){Churazov}, {Sazonov}, {Sunyaev}, {Forman},
  {Jones} \& {B{\"o}hringer}]{Churazov2005}
{Churazov} E., {Sazonov} S., {Sunyaev} R., {Forman} W., {Jones} C.,
  {B{\"o}hringer} H., 2005, \mnras, 363, L91

\bibitem[{Ciotti} \& {Ostriker}(2007)]{CiottiOstriker2007}
{Ciotti} L., {Ostriker} J.~P., 2007, \apj, 665, 1038

\bibitem[{Colberg} \& {di Matteo}(2008)]{Colberg2008}
{Colberg} J.~M., {di Matteo} T., 2008, \mnras, 387, 1163

\bibitem[{DeGraf} et~al.(2011){DeGraf}, {Di Matteo}, {Khandai}, {Croft},
  {Lopez} \& {Springel}]{DeGraf2011EarlyBHs}
{DeGraf} C., {Di Matteo} T., {Khandai} N., {Croft} R., {Lopez} J., {Springel}
  V., 2011, ArXiv e-prints

\bibitem[{DeGraf} et~al.(2010){DeGraf}, {Di Matteo} \& {Springel}]{DeGraf2010}
{DeGraf} C., {Di Matteo} T., {Springel} V., 2010, \mnras, 402, 1927

\bibitem[{Degraf} et~al.(2011){Degraf}, {Di Matteo} \&
  {Springel}]{DeGrafClustering2010}
{Degraf} C., {Di Matteo} T., {Springel} V., 2011, \mnras, 413, 1383

\bibitem[{Dekel} \& {Birnboim}(2006)]{DekelBirnboim2006}
{Dekel} A., {Birnboim} Y., 2006, \mnras, 368, 2

\bibitem[{Dekel} et~al.(2009){Dekel}, {Birnboim}, {Engel} et~al.]{Dekel2009}
{Dekel} A., et~al., 2009, Nature, 457, 451

\bibitem[{Di Matteo} et~al.(2008){Di Matteo}, {Colberg}, {Springel},
  {Hernquist} \& {Sijacki}]{DiMatteo2008}
{Di Matteo} T., {Colberg} J., {Springel} V., {Hernquist} L., {Sijacki} D.,
  2008, \apj, 676, 33

\bibitem[{Di Matteo} et~al.(2011){Di Matteo}, {Khandai}, {DeGraf}
  et~al.]{DiMatteo2011}
{Di Matteo} T., {Khandai} N., {DeGraf} C., {Feng} Y., {Croft} R., {Lopez} J.,
  {Springel} V., 2011, ApJL submitted

\bibitem[{Di Matteo} et~al.(2005){Di Matteo}, {Springel} \&
  {Hernquist}]{DiMatteo2005}
{Di Matteo} T., {Springel} V., {Hernquist} L., 2005, Nature, 433, 604

\bibitem[{Fan} et~al.(2006){Fan}, {Strauss}, {Becker} et~al.]{Fan2006}
{Fan} X., et~al., 2006, AJ, 132, 117

\bibitem[{Feng} et~al.(2011){Feng}, {Croft}, {Di Matteo} et~al.]{Feng2011}
{Feng} Y., et~al., 2011, ArXiv e-prints

\bibitem[{Ferrarese} \& {Merritt}(2000)]{FerrareseMerritt2000}
{Ferrarese} L., {Merritt} D., 2000, \apjl, 539, L9

\bibitem[{Gebhardt} et~al.(2000){Gebhardt}, {Bender}, {Bower}
  et~al.]{Gebhardt2000}
{Gebhardt} K., et~al., 2000, \apjl, 539, L13

\bibitem[{Graham} \& {Driver}(2007)]{GrahamDriver2007}
{Graham} A.~W., {Driver} S.~P., 2007, \apj, 655, 77

\bibitem[{Hopkins} et~al.(2007{\natexlab{a}}){Hopkins}, {Hernquist}, {Cox},
  {Robertson} \& {Krause}]{2007ApJ...669...45H}
{Hopkins} P.~F., {Hernquist} L., {Cox} T.~J., {Robertson} B., {Krause} E.,
  2007{\natexlab{a}}, \apj, 669, 45

\bibitem[{Hopkins} et~al.(2007{\natexlab{b}}){Hopkins}, {Richards} \&
  {Hernquist}]{Hopkins2007}
{Hopkins} P.~F., {Richards} G.~T., {Hernquist} L., 2007{\natexlab{b}}, \apj,
  654, 731

\bibitem[{Jiang} et~al.(2009){Jiang}, {Fan}, {Bian} et~al.]{Jiang2009}
{Jiang} L., et~al., 2009, AJ, 138, 305

\bibitem[{Johansson} et~al.(2008){Johansson}, {Naab} \&
  {Burkert}]{Johansson2008}
{Johansson} P.~H., {Naab} T., {Burkert} A., 2008, Astronomische Nachrichten,
  329, 956

\bibitem[{Kollmeier} et~al.(2006){Kollmeier}, {Onken}, {Kochanek}
  et~al.]{Kollmeier2006}
{Kollmeier} J.~A., et~al., 2006, \apj, 648, 128

\bibitem[{Kormendy} \& {Richstone}(1995)]{KormendyRichstone1995}
{Kormendy} J., {Richstone} D., 1995, ARA\&A, 33, 581

\bibitem[{Lemson} \& {Virgo Consortium}(2006)]{Lemson2006}
{Lemson} G., {Virgo Consortium} t., 2006, ArXiv Astrophysics e-prints: 0608019

\bibitem[Lopez et~al.(2011)Lopez, Degraf, DiMatteo, Fu, Fink \&
  Gibson]{Lopez2011}
Lopez J., Degraf C., DiMatteo T., Fu B., Fink E., Gibson G., 2011, in {
  Statistical and Scientific Databases Management Conference ({SSDBM})\/},
  Portland, OR

\bibitem[{Magorrian} et~al.(1998){Magorrian}, {Tremaine}, {Richstone}
  et~al.]{Magorrian1998}
{Magorrian} J., et~al., 1998, AJ, 115, 2285

\bibitem[{Mayer} et~al.(2007){Mayer}, {Kazantzidis}, {Madau}, {Colpi}, {Quinn}
  \& {Wadsley}]{Mayer2007}
{Mayer} L., {Kazantzidis} S., {Madau} P., {Colpi} M., {Quinn} T., {Wadsley} J.,
  2007, Science, 316, 1874

\bibitem[{Netzer} et~al.(2007){Netzer}, {Lira}, {Trakhtenbrot}, {Shemmer} \&
  {Cury}]{Netzer2007}
{Netzer} H., {Lira} P., {Trakhtenbrot} B., {Shemmer} O., {Cury} I., 2007, \apj,
  671, 1256

\bibitem[{Netzer} \& {Trakhtenbrot}(2007)]{NetzerTrakhtenbrot2007}
{Netzer} H., {Trakhtenbrot} B., 2007, \apj, 654, 754

\bibitem[{Sazonov} et~al.(2004){Sazonov}, {Ostriker} \& {Sunyaev}]{Sazonov2004}
{Sazonov} S.~Y., {Ostriker} J.~P., {Sunyaev} R.~A., 2004, \mnras, 347, 144

\bibitem[{Shakura} \& {Sunyaev}(1973)]{ShakuraSunyaev1973}
{Shakura} N.~I., {Sunyaev} R.~A., 1973, A\&A, 24, 337

\bibitem[{Shankar} et~al.(2009){Shankar}, {Weinberg} \&
  {Miralda-Escud{\'e}}]{Shankar2009}
{Shankar} F., {Weinberg} D.~H., {Miralda-Escud{\'e}} J., 2009, \apj, 690, 20

\bibitem[{Shankar} et~al.(2011){Shankar}, {Weinberg} \&
  {Miralda-Escude'}]{Shankar2011}
{Shankar} F., {Weinberg} D.~H., {Miralda-Escude'} J., 2011, ArXiv e-prints

\bibitem[{Shen} \& {Kelly}(2011)]{Shen2011}
{Shen} Y., {Kelly} B.~C., 2011, ArXiv e-prints

\bibitem[{Sijacki} et~al.(2007){Sijacki}, {Springel}, {di Matteo} \&
  {Hernquist}]{Sijacki2007}
{Sijacki} D., {Springel} V., {di Matteo} T., {Hernquist} L., 2007, \mnras, 380,
  877

\bibitem[{Sijacki} et~al.(2009){Sijacki}, {Springel} \&
  {Haehnelt}]{Sijacki2009}
{Sijacki} D., {Springel} V., {Haehnelt} M.~G., 2009, \mnras, 400, 100

\bibitem[{Springel}(2005)]{2005MNRAS.364.1105S}
{Springel} V., 2005, \mnras, 364, 1105

\bibitem[{Springel} et~al.(2005{\natexlab{a}}){Springel}, {Di Matteo} \&
  {Hernquist}]{SpringelFeedback2005}
{Springel} V., {Di Matteo} T., {Hernquist} L., 2005{\natexlab{a}}, \mnras, 361,
  776

\bibitem[{Springel} \& {Hernquist}(2003)]{SpringelHernquist2003}
{Springel} V., {Hernquist} L., 2003, \mnras, 339, 289

\bibitem[{Springel} et~al.(2005{\natexlab{b}}){Springel}, {White}, {Jenkins}
  et~al.]{Springel2005}
{Springel} V., et~al., 2005{\natexlab{b}}, Nature, 435, 629

\bibitem[{Trakhtenbrot} et~al.(2011){Trakhtenbrot}, {Netzer}, {Lira} \&
  {Shemmer}]{Trakhtenbrot2011}
{Trakhtenbrot} B., {Netzer} H., {Lira} P., {Shemmer} O., 2011, \apj, 730, 7

\bibitem[{Tremaine} et~al.(2002){Tremaine}, {Gebhardt}, {Bender}
  et~al.]{Tremaine2002}
{Tremaine} S., et~al., 2002, \apj, 574, 740

\bibitem[{Willott} et~al.(2010){Willott}, {Albert}, {Arzoumanian}
  et~al.]{Willott2010}
{Willott} C.~J., et~al., 2010, AJ, 140, 546

\bibitem[{Yoshida} et~al.(2006){Yoshida}, {Omukai}, {Hernquist} \&
  {Abel}]{Yoshida2006}
{Yoshida} N., {Omukai} K., {Hernquist} L., {Abel} T., 2006, \apj, 652, 6

\end{thebibliography}

\end{document}